\documentclass[aps,prl,twocolumn,groupedaddress,nofootinbib]{revtex4-2}
\usepackage{amsmath}
\usepackage{geometry}
\usepackage{setspace}
\usepackage{graphicx} 
\usepackage{hyperref}

\usepackage{fontawesome} 
\usepackage{orcidlink} 

\usepackage{booktabs,tabularray,tabularx,makecell} 
\usepackage{amssymb} 

\graphicspath{{./images/}}

\usepackage{xcolor}

\date{\today}
\usepackage{etoolbox}

\makeatletter
\patchcmd{\frontmatter@abstract@produce}
  {\vskip200\p@\@plus1fil
   \penalty-200\relax
   \vskip-200\p@\@plus-1fil}
  {}
  {}
  {}
\makeatother
\begin{document}
\singlespacing
\title{Frozen Surface Modes on a Moving Interface}

\author{S. Azar\,\orcidlink{0009-0002-9797-7485}}
\affiliation{Department of Physics and London Centre for Nanotechnology, King's College London, Strand, London WC2R 2LS, UK}
\author{M. J. Bhaseen\,\orcidlink{0000-0002-9089-2713}}
\affiliation{Department of Physics and London Centre for Nanotechnology, King's College London, Strand, London WC2R 2LS, UK}
\author{A. V. Zayats\,\orcidlink{0000-0003-0566-4087}}
\affiliation{Department of Physics and London Centre for Nanotechnology, King's College London, Strand, London WC2R 2LS, UK}
\author{F.~J. Rodr\'iguez-Fortu\~no\,\orcidlink{0000-0002-4555-1186}}
\email{francisco.rodriguez\_fortuno@kcl.ac.uk}
\affiliation{Department of Physics and London Centre for Nanotechnology, King's College London, Strand, London WC2R 2LS, UK}
\begin{abstract}
    Spatio-temporal modulation enables synthetic motion at effective velocities approaching the speed of light, providing new regimes for light–matter interaction. Traditional Cherenkov-type effects arise when the velocity of an emitter matches or exceeds the phase velocity of electromagnetic modes supported by a medium. Here, we study dispersive systems in which phase and group velocities differ markedly. Specifically, we explore the case of group-velocity matching for surface waves, where the emitter moves at the same velocity as the flow of energy. This gives rise to frozen surface modes which are stationary in the emitter frame, accompanied by resonant energy accumulation. The result is a dramatic increase of the local density of optical states, the power extracted from the emitter, and the optomechanical forces and torques it experiences. Since surface modes naturally exhibit slow group velocities, this is accessible at lower relative speeds than phase-velocity effects. This phenomenon provides a route to enhanced light-matter interaction via real or synthetic motion.
    
\end{abstract}

\maketitle

Rapid progress in time-varying photonics has revealed a wide range of electromagnetic phenomena enabled by temporal modulation of matter \cite{Engheta20204dmetamaterial}. Pioneering experiments have demonstrated temporal analogues of ubiquitous spatial phenomena, including reflection and refraction, double slit interference, time-crystals and momentum-space bandgaps
\cite{Xiao2014temporal_boundary,Moussa2023time_boundary_experiment,Jones2024time_boundary_experiment,Tirole2023time_double_slit,paco2023time_double_slit,Akkermans2012time_slit,Zurita-Snchez2009temporal_band_gap, Shaltout2016temporal_band_gap, Martnez-Romero2016temporal_band_gap,Wang2023time_crystal,Liu2023time_crystal}. Potential applications include non-reciprocal wave propagation \cite{Shen2016non_reciprocity,Vezzoli2018time_reversal}, amplification \cite{Pan2023amplification,Xiong2025amplification}, and “synthetic” motion at luminal or superluminal speeds \cite{Philbin2008, Huidobro2019, Galiffi2022, Harwood2025}. 

The recent developments in synthetic motion make it timely to explore the light-matter interaction of emitters moving in, or in proximity to, a medium.
In the classic example of a point charge moving through a homogeneous bulk medium, Cherenkov radiation arises when the speed of the charge exceeds the phase-velocity \(v \ge v_p\) \cite{Jackson1998, Cherenkov1937, Smith1953, Ginzburg1996}. 
However, in many optical and near-field settings the relevant source is electrically neutral and is well described by an oscillating emitter with rest-frequency \(\omega_0\). As we will show, this temporal degree of freedom opens the possibility of group-velocity matching, which does not happen for point-charges.
Possible sources include quantum emitters, classical scatterers, and the fluctuating dipoles in thermal emission and the Casimir effect. For work on the near-field effects of charges and emitters moving near interfaces at the phase velocity threshold see Refs \cite{rullhusen1998novel, Denton1998moving_charge, Jackson1998, tamm1937moving_electrons, Ginzburg1996, Kobzev2014overview_moving_charge, Liu2012cherenkov_moving_electron,horsley2012canonical,Rubino2012negative_frequency}.

In this work, we study an emitter with rest frequency $\omega_0$ in relative motion with a planar interface that supports dispersive surface modes. We show that a distinct regime arises when the relative speed matches the group velocity \(v_g\) of a supported mode. The excited surface mode becomes stationary---or frozen---in the emitter frame. This produces a strong increase in the local density of optical states (LDOS), as well as enhanced extracted power and optomechanical forces and torques. This group-velocity condition is inherently different from the more familiar phase-velocity (Cherenkov-type) threshold, and can be reached at lower speeds. In contrast to bulk waves, where $v_p$ and $v_g$ generally differ only weakly through material dispersion, surface modes can exhibit strong modal dispersion with $v_g \ll v_p$. While this mechanism is generic to any dispersive surface mode, we explore this via surface plasmon polaritons (SPPs), the collective charge–photon excitations at metal–dielectric interfaces. This serves as an illustrative platform with a strongly reduced group velocity \cite{Maier2007plasmonic}.

\begin{figure*}[ht]
    \centering
\includegraphics[width = 17cm]{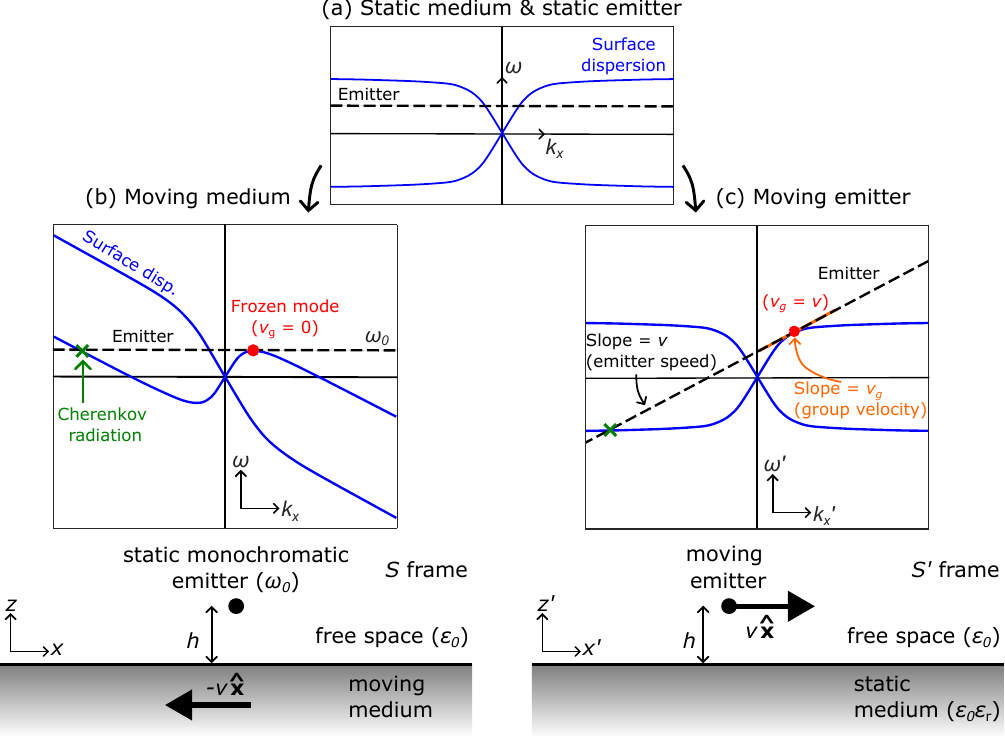}
    \caption{Dispersion relation of surface plasmon excitation (solid) and emitter spectrum (dashed) in (a) the static case where the emitter and interface are stationary; (b) in the emitter reference frame $S$ with stationary emitter and moving interface; (c) in the surface reference frame $S'$ with stationary interface and moving emitter. The red dots indicate the group-velocity matching condition (frozen mode) and the green crosses indicate the phase-velocity matching condition (Cherenkov radiation).}
    \label{fig:moving_dipole}
\end{figure*}

\textit{Setup}.--- We consider a planar interface at $z=0$ between vacuum and a dispersive medium with rest-frame permittivity $\varepsilon_r(\omega)$, as shown in Fig.~\ref{fig:moving_dipole}. A monochromatic point dipole, oscillating at frequency $\omega_0$, is placed at distance $z=h$ above the surface. Although the formalism applies for arbitrary $h$, we focus on subwavelength separations to enable near-field coupling to surface modes. The emitter and the surface are in relative motion parallel to the surface with speed $v$. In this translationally invariant geometry, both frequency and in-plane wave vector are conserved, so surface-mode excitation is only possible when the emitter spectrum intersects the surface-mode dispersion in Fourier space \cite{paco2019guided_mode}. In the static case ($v=0$), this reduces to the familiar condition that the dispersion $\omega(k)$ crosses $\omega_0$ as shown in Fig.~\ref{fig:moving_dipole}(a).

We study the relative motion of the emitter along the interface using a Green's function formulation with Lorentz-transformed spectral components to account for reflection from the moving interface. Early works on moving surface reflection considered plane waves \cite{Yeh1965moving_surface, Pyati1967moving_surface, Shiozawa1967moving_surface, Lee1967moving_surface, Shiozawa1968moving_surface, Mukherjee1973moving_surface}, but for near-field emitters one needs to incorporate evanescent waves \cite{azar2025reflectioncoefficients}. Relative motion tilts the Fourier-space spectra; Fig.~\ref{fig:moving_dipole}. In the rest frame $S$ of the emitter, its spectrum is unaffected ($\omega=\omega_0$), while the moving surface tilts the surface-mode dispersion; Fig.~\ref{fig:moving_dipole}(b). In the rest frame $S'$ of the surface, the surface-mode dispersion is that of an interface at rest \cite{Maier2007plasmonic}, but the emitter spectrum is Doppler-shifted. This corresponds to the dashed line of slope $v$ in the $(k_x',\omega')$ plane in Fig.~\ref{fig:moving_dipole}(c). Table~\ref{tab:key_equations_frames} summarises the corresponding Fourier-space relations in both frames for SPP modes.

\textit{Crossing point}.--- The tilt of the spectra gives rise to new crossing points between the emitter and the surface dispersion relation that were previously inaccessible. The green cross in Figs~\ref{fig:moving_dipole}(b,c) corresponds to the more usual phase-velocity matching. In the surface frame $S'$ the emitter outruns the SPP phase velocity and can launch modes that are otherwise kinematically forbidden \cite{Liu2012cherenkov_moving_electron,Jackson1998}. This regime has been studied and linked to vacuum energy extraction \cite{horsley2012canonical} and negative-frequency resonant radiation \cite{Rubino2012negative_frequency}. 
Cherenkov radiation is usually formulated for a moving point charge, which corresponds to an emitter with $\omega_0=0$. Therefore, the phase-matching condition occurs at the $\omega=0$ axis. An oscillating emitter shifts the phase-matching condition to $\omega_0>0$, as shown in Fig.~\ref{fig:moving_dipole}(b). This matching can always be achieved for any non-zero speed, but happens at very large $|k_x|$ values. In practice, material absorption and near-field decay with distance limit the achievable $|k_x|$ \cite{Maier2007plasmonic}.

In contrast, a more accessible regime related to the surface mode group velocity can be found at lower $|k_x|$ values. For a suitably tuned emitter frequency $\omega_0$, the red dot in Figs.~\ref{fig:moving_dipole}(b,c) shows the presence of a group velocity matching condition. In the rest frame $S$ of the emitter, the SPP dispersion exhibits a stationary point corresponding to zero group velocity, Fig.~\ref{fig:moving_dipole}(b), which we refer to as a frozen mode. Equivalently, in the surface frame, the emitter velocity matches the SPP group velocity, so that the tilted emitter spectrum is tangent to the dispersion curve; Fig.~\ref{fig:moving_dipole}(c). This condition is experimentally easier to attain than Cherenkov-type excitation because dispersive surface modes can have very small group velocities. For surface plasmon polaritons, $v_g \ll v_p$, especially near the surface plasmon frequency $\omega_{\mathrm{sp}}$ \cite{Maier2007plasmonic}.

\begin{table*}
  \caption{Fourier-space relations for the emitter spectrum and SPP dispersion in the emitter frame $S$ and the surface frame $S'$. The frequencies and wave-vectors are related by the Lorentz transformations $\omega'=\gamma(\omega+\beta c k_x)$, $k_x'=\gamma(k_x+\beta\,\omega/c)$, $k_y'=k_y$, where $\beta=v/c$ and $\gamma=(1-\beta^2)^{-1/2}$. Here $k_t=\sqrt{k_x^2+k_y^2}$. The material response $\varepsilon_r(\omega')$ is evaluated at the surface-frame frequency $\omega'$ with the medium at rest. In frame $S$, the appearance of $\omega'$ in $\varepsilon_r$ produces the tilt shown in Fig.~\ref{fig:moving_dipole}(b); because $\omega'$ depends on $k_x$, the SPP dispersion relation is an implicit equation which is solved self-consistently.}
  \label{tab:key_equations_frames}
  \centering
  \begin{tblr}{
    width = \linewidth,
    colspec = {Q[l,m] X[c,m] X[c,m]},
    rowsep = 4pt,
    hline{1} = {1-3}{solid},
    hline{2} = {1-3}{solid},
    hline{4} = {1-3}{solid},
    hline{5} = {1-3}{solid},
  }
    & \textbf{Emitter reference frame }$(S)$
    & \textbf{Surface reference frame }$(S')$ \\
   \makecell[c] {\textbf{Emitter}} &
    $\displaystyle \omega = \omega_0 \quad\text{(constant)}$ &
    $\displaystyle \omega'=\frac{\omega_0}{\gamma} + v\,k_x'\quad\text{(tilted)}$ \\
    \SetCell[r=1]{l,m} \makecell[c]{\textbf{SPP}\\\textbf{dispersion}}  &
    $\displaystyle k_{x} = \frac{\pm \sqrt{\varepsilon_r(\omega')} -\beta \sqrt{\varepsilon_r(\omega')+1}}
       {\sqrt{\varepsilon_r(\omega')+1} \mp \beta \sqrt{\varepsilon_r(\omega')}}\;
      \frac{\omega}{c} \,(\text{}*)$ &
    $\displaystyle k_{t}^{\prime} = \frac{\omega'}{c}\,
      \sqrt{\frac{\varepsilon_r(\omega')}{1+\varepsilon_r(\omega')}}\quad\text{(static surface)}$ \\
    \SetCell[c=3]{c} \footnotesize $^{*}$ For simplicity, the tilted SPP dispersion in frame $S$ is given for $k_y=0$ \cite{azar2025reflectioncoefficients}. \\
  \end{tblr}
\end{table*}

\textit{Matching condition}.--- Coupling to the frozen-mode requires a matching condition between the emitter frequency $\omega_0$ and the relative speed $v$, which is not attainable for point charges with $\omega_0=0$. Setting $v_g=\partial\omega/\partial k_x=0$ for the moving-surface dispersion in the emitter frame $S$ (Table~\ref{tab:key_equations_frames}) yields the condition 
\begin{equation}
\label{eq:frozen mode}
    k_\text{fsm}' - \beta\,\frac{\omega_\text{fsm}'}{c}\, \frac{\left[\varepsilon_r(\omega_\text{fsm}')\right]^2+1}{\left[\varepsilon_r(\omega_\text{fsm}')+1\right]^2}=0,
\end{equation}
where $\omega_\text{fsm}'$ and $k_\text{fsm}'$ are the surface-frame frequency and in-plane wave vector of the frozen surface mode (fsm), and $\beta=v/c$. The primed and unprimed quantities are related by a Lorentz transformation, so that Eq.~(\ref{eq:frozen mode}) must be solved self-consistently with the SPP dispersion relation in Table~\ref{tab:key_equations_frames}. Eq.~(\ref{eq:frozen mode}) can be interpreted geometrically as a tangency condition between the emitter spectrum and the SPP dispersion relation (red dots in Fig.~\ref{fig:moving_dipole}). 

\begin{figure*}[ht]
    \centering
    \includegraphics[width = 17cm]{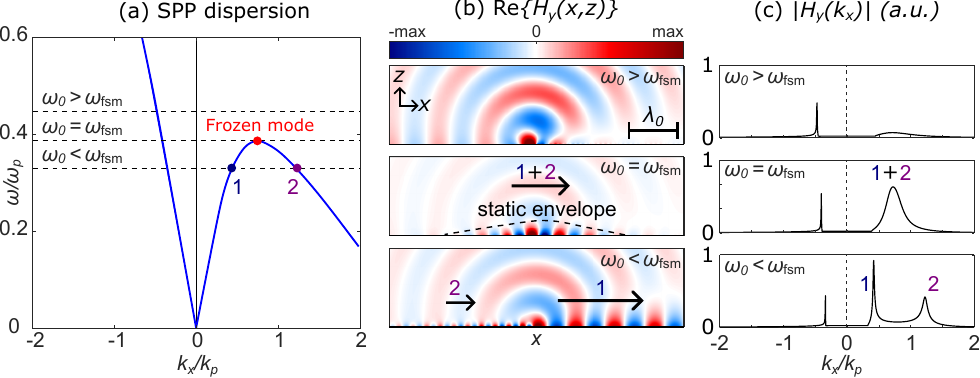}
    \caption{
    Regimes of mode excitation for  an $x$-polarized line dipole placed at $h=0.05\lambda_0$ above a moving surface, where $\lambda_0=2\pi c/\omega_0$. (a) Tilted dispersion relation of a lossless Drude surface with $v=0.25c$, corresponding to Fig.~\ref{fig:moving_dipole}(b). The dashed lines correspond to (i) $\omega_0<\omega_{\text{fsm}}$, (ii) $\omega_0=\omega_{\text{fsm}}$ and (iii) $\omega_0>\omega_{\text{fsm}}$, where $\omega_{\text{fsm}}$ is the frequency of the frozen surface mode in Eq.~\ref{eq:frozen mode}. (b) Magnetic field distribution for the three cases in panel (a), where the arrows represent the phase velocity of each mode. Mode 2 is a backwards propagating mode, emitted to the left with its phase fronts moving to the right. (c) Spatial spectrum of the magnetic field shown in panel (b) at $z=0$. Modes 1 and 2 coalesce at $\omega_0=\omega_\text{fsm}$ corresponding to the frozen mode with a broadened spectrum.}
    \label{fig:merging_modes}
\end{figure*}

\textit{Nature of the frozen mode}.--- It is instructive to consider the emergence of the red crossing point in Figs.~\ref{fig:moving_dipole}(b,c) as the value of $\omega_0$ is changed for a fixed $v$. The appearance of the frozen mode can be interpreted as the coalescence of two SPP modes as $\omega_0\to\omega_{\text{fsm}}$. 
Fig.~\ref{fig:merging_modes} compares three parameter choices, below ($\omega_0<\omega_{\text{fsm}}$), at ($\omega_0=\omega_{\text{fsm}}$), and above ($\omega_0>\omega_{\text{fsm}}$) the frozen-mode condition in Eq.~\ref{eq:frozen mode}.

We start by considering the rest frame $S$ of the emitter. Below the frozen-mode condition, two SPP crossing points can be seen for $k_x>0$, labelled as modes~1 and~2 in Fig.~\ref{fig:merging_modes}(a). Mode~1 is the right-propagating SPP that would exist even if $v=0$. Mode~2 is enabled by the motion of the surface, corresponding to Cherenkov radiation in the limit $\omega_0=0$. It is a so-called backward-wave \cite{Waldron1964backward_wave,Tremain2018backward_mode} with opposite phase ($v_p>0$) and group ($v_g<0$) velocities. At the frozen-mode condition, modes 1 and 2 merge in $k_x$. In real space, the electromagnetic field forms a wavepacket pinned to the emitter, while phase fronts slide within the stationary envelope; Fig.~\ref{fig:merging_modes}(b). This coalescence also appears as a pronounced broadening of the SPP spectrum in momentum space; Fig.~\ref{fig:merging_modes}(c). As we will discuss below, this broadening leads to large increases of the LDOS, the power extracted from the emitter, and the optomechanical forces and torques. In a lossless model, all of these quantities diverge due to a pile-up of the surface modes. This divergence is analogous to a Van Hove singularity in electronic systems \cite{VanHove1953}. In practice, absorption regularizes the response, and the EM field calculations in Fig.~\ref{fig:merging_modes}(b) include a small loss. Above the frozen-mode condition, the two SPP solutions annihilate and no crossing point remains. This leads to an abrupt drop in the LDOS and related observables, causing sign reversals in forces and torques, as shown below.

These distinct regimes can also be understood in the rest frame $S'$ of the surface. Below the frozen mode condition, the emitter is moving slower than the phase velocity of modes 1 and 2, but faster than the group velocity of mode 2. At the frozen-mode condition, the two modes coalesce into a mode whose group velocity matches the emitter. As such, energy cannot outrun the emitter and it accumulates locally. Above the frozen-mode condition, the Doppler shift in the emitter frequency pushes the emission above $\omega_{\mathrm{sp}}$, shutting off SPP excitation. 

\begin{figure}[hb]
    \includegraphics[width = 8.5cm]{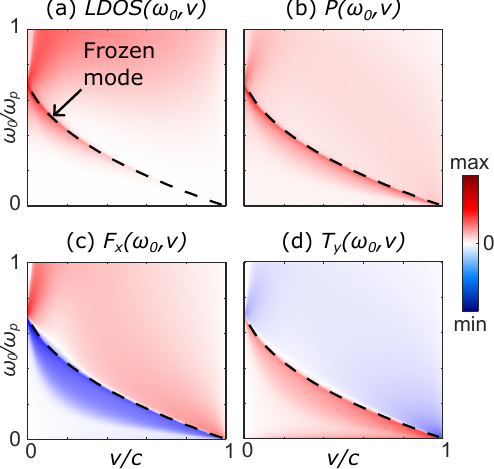}
    \caption{Color map of (a) LDOS, (b) power extracted, (c) lateral optical force, and (d) torque experienced by an x-polarized line dipole, as a function of $\omega_0$ and $v$, for $h = 0.05\lambda_0$. The dipole moment is normalised to emit unit power in free space. The dashed lines correspond to the frozen mode condition, where enhancements in the response are observed. The effect of small losses are included in the Drude model for $\varepsilon_r$, which regularizes the response. The results are plotted using a signed logarithmic color scale to show the sign reversal in panels (c) and (d).}
    \label{fig:enhancement_figure}
\end{figure}

\textit{Physical consequences}.--- In Fig.~\ref{fig:enhancement_figure}(a) we show the increase of the $\mathrm{LDOS}\propto \mathrm{Im}\,\mathrm{Tr}\,\overleftrightarrow{\mathbf{G}}$ \cite{economou2006green_LDOS,Barnes2020LDOS} at the frozen mode condition, corresponding to the dashed curve. Here, $\overleftrightarrow{\mathbf{G}}$ is the dyadic Green's function of the moving surface. 
At zero relative speed $v=0$, the frozen mode condition happens when $\omega_0=\omega_p/\sqrt{2}$, where $v_g\to0$ for SPPs. This corresponds to the well known LDOS increase in the proximity of a stationary plasmonic surface \cite{Ford1984LDOS_increase}. For non-vanishing speeds $v\neq0$, the condition on $\omega_0$ red-shifts, and frozen mode SPP excitation occurs at more accessible $|k_x|$ values. 
As shown in panels Fig.~\ref{fig:enhancement_figure}(b-d), the increase in LDOS leads to corresponding enhancements in the extracted power $P$ and the optomechanical force $\mathbf{F}$ and torque $\mathbf{T}$. 
The appearance of a lateral force and optical torque relies on the broken left-right symmetry induced by the motion. This biases SPP excitation even for symmetric emitters. The conservation of linear and angular momentum leads to a lateral recoil force and torque \cite{paco2015lateral_force,paco2018current_induced_unidirectional_SPP,paco2019lateral_force,Manjavacas2017casimir,Wang2014lateral_force,Canaguier-Durand2013torque,paco_sebastian2024radiation_forces_torques_optics}.

A notable feature in Fig.~\ref{fig:enhancement_figure} is an asymmetry in the behavior below and above the frozen mode. 
The response varies smoothly below the dashed line, but drops abruptly above it. This sharp decrease reflects the disappearance of modes 1 and 2. As can be seen in Figs.~\ref{fig:enhancement_figure}(c,d), this also leads to a reversal in the sign of the force and torque. 
Approaches based on single-photon emission events near moving surfaces \cite{horsley2012canonical} do not naturally capture these effects which rely on coherent photon accumulation in the near field. 

\begin{figure*}[ht]
    \centering
    \includegraphics[width = 17cm]{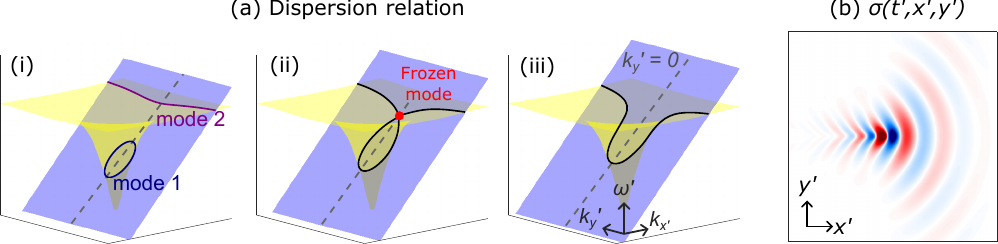}
    \caption{(a) Intersection of the Doppler shifted emitter spectrum (lilac plane) with the three-dimensional SPP dispersion (yellow) in $(k_x',k_y',\omega')$ space, corresponding to the surface frame $S'$. Their intersections (solid lines) result in the excited SPPs (i) below, (ii) at, and (iii) above the frozen-mode condition. The dashed line corresponds to $k_y'=0$, representing the 2D results. The modes coalesce at the frozen mode condition, where group-velocity matching is achieved (red dot). (b) Surface charge density $\sigma(t',x',y')$ at $z'=0$ and $t'=0$ at the frozen-mode condition. The frozen wave-packet is accompanied by conical and cylindrical waves, reflecting the presence of two segments of the dispersion relation in panel (a)(ii). An interactive visualization corresponding to panel~(a) is available at \href{https://frarodfo.github.io/moving-interface-frozen-modes/}{\faGithub~frarodfo.github.io/moving-interface-frozen-modes/} and archived in Ref.~\cite{rodriguez_fortuno_2026_mifm}.}
    \label{fig:3D_dispersion}
\end{figure*}

\textit{Three-dimensional case}.--- Having explored the two-dimensional case, we turn our attention to three dimensions. In the rest frame $S'$ of the surface, the SPP dispersion forms a sheet in $(k_x',k_y',\omega')$, while the moving monochromatic emitter yields a tilted ``emission plane''; Fig.~\ref{fig:3D_dispersion}(a). Below the frozen-mode condition, their intersection consists of two disjoint contours corresponding to modes 1 and 2; Fig.~\ref{fig:3D_dispersion}(a)(i). At the frozen-mode condition the contours touch at a single point of  group-velocity matching. As in the 2D case, this can be interpreted as a tangency between the emission plane and the dispersion sheet; Fig.~\ref{fig:3D_dispersion}(a)(ii). Above this threshold, a single intersection curve remains; Fig.~\ref{fig:3D_dispersion}(a)(iii). The 2D result is recovered via the dashed line corresponding to $k_y'=0$.

The contours of intersection depicted by the solid lines in Fig.~\ref{fig:3D_dispersion}(a) determine the excitations in the system. These have a direct signature in the real space EM field and surface charge distribution. Fig.~\ref{fig:3D_dispersion}(b) shows the surface charge density at $z=0$ at the frozen-mode condition. The profile consists of a combination of cylindrical wavefronts and Cherenkov-like cones, corresponding to different segments of the contours. The cylindrical wavefronts are associated to the closed loop in Fig.~\ref{fig:3D_dispersion}(a)(ii), where the group velocity outruns the emitter. The conical wavefronts correspond to the open branches, where surface waves with slow group velocity lag behind.

\textit{Conclusion}.--- We have identified a route to enhanced light-matter interaction via motion. In particular, we found a near-field regime in which relative motion between an emitter and a dispersive interface matches the group velocity of a supported surface mode. This creates a frozen wavepacket in the emitter frame which strongly increases the LDOS, extracted power, and optomechanical forces and torques. Unlike the more familiar phase-velocity (Cherenkov-type) condition, group-velocity matching leads to resonant energy accumulation which can dominate the response. It would be interesting to explore this behavior in space-time modulated metamaterials which can emulate luminal or superluminal synthetic motion. More broadly, the same frozen-mode enhancement mechanism should amplify the response for fluctuation-induced dipoles. This suggests opportunities for observing and controlling Casimir forces and torques near moving or synthetically moving interfaces. This effect has broad relevance to all wave domains, including electromagnetic, elastic and acoustic phenomena.

\section*{Acknowledgments}
S. A. is supported by EPSRC through an NMESFS-2024 scholarship. M.J. B., A.V. Z. and F.J. R.-F. were supported by the Engineering and Physical Sciences Research Council (EPSRC) grant META4D (EP/Y015673).

\bibliography{reference.bib}

\end{document}